\documentclass[11pt]{aastex}
\usepackage{emulateapj5,apjfonts}
\usepackage{onecolfloat}
\usepackage{epsf}

\slugcomment{{\sc Accepted by ApJ:} 16th October 2000}

\newcommand{\ha}{{\rm H$\alpha$ }}
\newcommand{\hb}{{\rm H$\beta$ }}
\newcommand{\hahb}{{\rm H$\alpha$/H$\beta$ }}
\newcommand{\hi}{{\rm H{\sc i} }}
\newcommand{\hans}{{\rm H$\alpha$}}
\newcommand{\hii}{{\rm H\,{\sc ii} }}
\newcommand{\fir}{{\rm FIR }}
\newcommand{\firns}{{\rm FIR}}
\newcommand{\fuv}{{\rm FUV }}
\newcommand{\fuvns}{{\rm FUV}}
\newcommand{\nuv}{{\rm NUV }}
\newcommand{\niil}{{\rm $[$N\,{\sc ii}$]\,\lambda$6583 }}
\newcommand{\nii}{{\rm $[$N\,{\sc ii}$]$ }}
\newcommand{\msun}{${\rm M_{\sun}}$}
\newcommand{\firfuv}{{\rm FIR/FUV }}
\newcommand{\firfuvns}{{\rm FIR/FUV}}

\shorttitle{Comparing Star Formation Rate Indicators}
\shortauthors{Bell \& Kennicutt}

\begin{document}


\def\head{

\title{A Comparison of UIT Far-Ultraviolet and \ha Star Formation Rates}
\author{Eric F. Bell and Robert C. Kennicutt, Jr.}
\affil{Steward Observatory, University of Arizona, 933 N. Cherry Ave., 
	Tucson, AZ 85721, USA}
\email{ebell,robk@as.arizona.edu}

\begin{abstract}
We have used archival ultraviolet (UV) imaging of 50 nearby star-forming
galaxies obtained with the Ultraviolet
Imaging Telescope (UIT) to derive integrated near-UV and far-UV 
magnitudes, and have combined these data with 
\hans, far-infrared, and thermal radio continuum measurements to explore
the consistency of UV and \ha star formation rates (SFRs).
In agreement with previous studies, we find that the UV and \ha SFRs
are qualitatively consistent, even before corrections for extinction
are applied.  The uncorrected UV SFRs are systematically lower
by a factor of 1.5 (with a factor of two scatter) 
among luminous galaxies with SFR 
$\ga 1$ \msun$\,{\rm yr}^{-1}$, indicating a higher effective 
attenuation of the far-UV radiation.  
Among less luminous galaxies there is no significant 
offset between the \ha and far-UV SFR scales.
This behavior is consistent with 
those of higher-redshift samples observed by Sullivan
et al., Glazebrook et al., and Yan et al. for comparable ranges of 
galaxy luminosities and absolute SFRs.  Far-infrared and thermal
radio continuum data available for a subset of our sample allow us to
estimate the attenuation in the UV and at \ha independently.  The UV and
\ha attenuations appear to be correlated, and confirm systematically
higher attenuations in the UV.  Although the galaxies in our  
sample show modest levels of attenuation (with median values of 0.9 mag
at \ha and 1.4 mag at 1550 \AA), 
the range across the sample
is large, $\sim$4 mag for \ha and $\ga$5 mag in the far-UV (1550 \AA).
This indicates that the application of a single characteristic extinction
correction to \ha or UV SFRs is only realistic for large, well-defined
and well-studied galaxy samples, 
and that extinction bias may be important for UV or
emission-line selected samples of star-forming galaxies.
\end{abstract}

\keywords{galaxies : general --- galaxies : photometry --- galaxies : stellar
	content --- galaxies : evolution --- ultraviolet : galaxies ---
	dust, extinction}
}

\twocolumn[\head]

\section{INTRODUCTION}

Until recently most of our information about the systematic behavior
of star formation rates (SFRs) in normal nearby galaxies has been 
based on measurements of the \ha emission line 
\citep[e.g.][]{kk83,gallego95}.  In contrast, 
SFRs for $z > 1$ galaxies are primarily based on 
observations of the redshifted ultraviolet 
(UV) continuum \citep[e.g.][]{madau98,steidel99}.  Comparable UV observations
for nearby galaxies are being accumulated 
\citep[e.g.][]{donas87,tresse98,buat99,sullivan00}, and these have
made it possible to derive the cosmic evolution of the volume-averaged
SFR in a self-consistent manner.  However, questions remain about the
systematic accuracy of the UV-derived SFRs, both in terms of the absolute
SFR scale and possible redshift-dependent biases in that scale.
Similar questions also apply to the \hans-based SFRs, and with the 
extension of this technique to high-redshift galaxies by several groups
\citep[e.g.][]{glaze99,yan99,moorcroft00}
it is important to understand the limitations of this technique also.

Although \ha and UV measurements are now available for samples of hundreds
of galaxies in each case, there have been only a few direct comparisons
of the respective SFRs for galaxies in common.  Comparisons of three samples
of high-redshift galaxies ($z = 0.7 - 2.2$) by \citet{glaze99},  
\citet{yan99}, and \citet{moorcroft00} 
show that the uncorrected \ha fluxes yield SFRs that are
$\sim$3 times higher than derived from the 2800 \AA\ UV flux.  The
simplest explanation for this difference would be a systematically
higher extinction in the UV.  However, similar comparisons using nearby
samples present a less clear picture.  \citet{buat87} and \citet{buat92}
compared balloon-based UV and ground-based \ha measurements
of small samples of galaxies and found a general consistency between
the extinction-corrected SFR scales.  Furthermore, \citet{buat96}
estimated UV extinctions for a large sample of star-forming spirals
and estimated typical extinctions that are comparable to or lower
than those at \hans.  The most comprehensive comparison to 
date of UV and \hans-based 
SFRs was carried out by \citet{sullivan00}, 
who analyzed balloon-based UV (2000 \AA) and \ha fluxes for 273 galaxies in the
redshift range $0 < z < 0.5$.  The galaxies in this sample span
nearly 4 orders of magnitude in SFR, ranging from dwarfs to massive spirals;
many galaxies are undergoing intense star formation bursts.  Although they
observed a strong correlation between UV and \hans-derived SFRs, they
also found that the UV/\ha flux ratio decreases significantly with
increasing SFR.  In low-SFR systems the UV fluxes
consistently yield {\it higher} SFRs than \hans, while the reverse is true
for galaxies with the highest SFRs.  They attribute most of the 
differences to rapid variability in the SFR over the lifetimes of 
the starbursts.

These results underscore the need for a complimentary comparison
of UV and \ha SFRs for a sample of well-resolved nearby galaxies
with a fuller range of star formation histories.  
In this paper we primarily compare far-UV (\fuvns) and 
\ha SFRs for a sample of 50 nearby star-forming galaxies imaged 
using the Ultraviolet Imaging Telescope \citep[UIT;][]{stecher97}.
This sample is much smaller than that studied by \citet{sullivan00}
but it has several advantages which complement the strengths of their
study.  The UIT sample is far from complete: rather, it 
was selected to span a full range of morphological
types, luminosities, SFRs, and star formation properties, with the
bulk of the sample composed of high-luminosity spiral galaxies.
This virtually eliminates the problem of rapidly variable SFRs that  
are important when analyzing starburst galaxy samples, and allows us to compare
the zeropoints of the UV and \hans-based SFR scales on a self-consistent
basis.  Another advantage of this sample is the availability of truly
integrated UV and \ha fluxes, which eliminates any possible effects
of aperture undersampling at \hans.  

This paper concentrates on a comparison of the integrated fluxes and
SFRs of the UIT/\ha sample.  Our main objectives are to: 
1) compare the SFRs derived from \fuv and \ha luminosities, with and
without corrections for dust attenuation;  2) compare the behavior of our
UV/\ha SFRs with those derived for other samples at low and 
high redshift; and 3) 
obtain insight into the amount of attenuation suffered
by \fuv and \ha in this sample of star-forming galaxies.
Future papers will compare the UV and emission-line properties of 
individual star-forming regions in these galaxies.

In this paper we follow the convention of \citet{fluxratio}
in using the term `attenuation'
to refer to the net loss of radiation from a galaxy at a given
wavelength.  For a point source such as a star this would be
the equivalent to the extinction (itself a combination of dust
absorption and scattering), but for a galaxy the radiation reaching
the observer is the result of a complicated radiative transfer
across multiple lines of sight with enormous variations in 
local optical depth.  In the specific context of this paper
it will generally refer to the correction factor that must be
applied to the observed integrated UV or \ha flux to derive
the actual SFR due to the effects of interstellar dust (and 
exclusive of other physical effects such as
escape of ionizing photons from a galaxy, etc.).

The remainder of the paper is organized as follows.
In \S \ref{sec:data}, we present the data: near-UV (NUV) and 
\fuv magnitudes, \ha line luminosities, far-infrared (FIR) luminosities and 
thermal radio continuum fluxes at 1.4 GHz.  We also present
the SFR calibrations for \ha and \fuv luminosities.
In \S \ref{sec:sfr}, we compare the raw \ha and \fuv
SFRs, and interpret the discrepancies in terms of variations in
star formation histories and differences in dust attenuation.
In \S \ref{sec:atten}, we examine the role of attenuation in 
more detail.
Finally, in \S \ref{sec:conc}, we present our conclusions.

\section{DATA} \label{sec:data}

Our analysis is primarily based on a sample of 50 nearby star-forming 
galaxies imaged in the \fuv
by the UIT.
The UIT provides well-resolved
(FWHM $\sim$ 3$\arcsec$), wide-field (field radius $\sim$ 40$\arcmin$)
images in both the \nuv and \fuvns.
This spatial resolution is useful, as it allows us to evaluate the 
effects of local
processes in determining the global \fuv fluxes 
\citep[e.g.][]{marcum00,kuchinski00}.  

\subsection{FUV Data}

The \fuv data was taken from the UIT archive at the Multimission 
Archive at the Space Telescope Science Institute.  As part of 
the Astro payload, the 38-cm UIT flew on two Space Shuttle missions in 1990 Dec
and 1995 Mar.  For the purposes of this analysis, we selected 
star-forming spiral and irregular galaxies with sizes comparable to
or smaller than the UIT field size (therefore excluding the Large and 
Small Magellanic Clouds, M31, and M33 from our sample).  We also
excluded from our analysis E/S0 galaxies with no evidence of massive
star formation (the UV emission in such objects is thought to arise
from evolved stellar populations), and we excluded galaxies for which
the bulk of the \ha emission arises from a Seyfert or LINER nucleus 
(e.g. NGC~2992, NGC~3227, NGC~4151).  Two other galaxies, 
NGC~1268 and UGC~2665 were omitted from the sample because 
they were not detected in the \fuv and no 
H$\alpha$, far-infrared (\firns) or radio fluxes are available in any case.   
Most of the sample galaxies are luminous later-type spirals: the remaining 
galaxies are split between earlier type spirals, irregulars and 
starbursting galaxies.  The starbursts in this sample are relatively faint
compared to those of e.g.\ \citet{calzetti94}.

The properties of the galaxies in our sample are summarized in Table 
\ref{tab:sample}.  Listed are the galaxy name, coordinates, type, distance, 
distance reference, and 
Galactic foreground extinction in the $V$ band, 
as estimated by Schlegel, Finkbeiner,
\& Davis \markcite{sfd}(1998).  
When a direct distance
determination was not available it was derived using 
H$_0 = 75$\,km\,s$^{-1}$\,Mpc$^{-1}$.  Note that most of the 
conclusions of this paper are distance independent as they rely
only on flux ratios: only the absolute magnitude and SFR axes 
in Fig.\ \ref{fig:sfi} are affected by distance uncertainties.

Every galaxy was imaged in at least one of the two wide \fuv passbands 
used by the UIT:
passband B1 
($\lambda_{\rm eff} \sim 1521$\,\AA; $\Delta\lambda \sim 354$\,\AA)
and passband B5 
($\lambda_{\rm eff} \sim 1615$\,\AA; $\Delta\lambda \sim 225$\,\AA).  The
latter filter was used during daylight observations to exclude dayglow
emission lines \citep{waller95}.  Additional observations
in the \nuv from the Astro-1 mission were also analyzed when available.
These were made in passband A1 
($\lambda_{\rm eff} \sim 2488$\,\AA; $\Delta\lambda \sim 1147$\,\AA).
Table \ref{tab:exp} lists the filters and exposure times for the
specific observations that were analyzed in this paper.
We usually chose the longest exposure images with the 
best signal-to-noise, with exposure times
ranging between $\sim$250\,s and $\sim$1500\,s.  

The UIT imaged galaxies using an image intensifier, recording 
the images on photographic film.  The images have 
been linearized, flat-fielded, flux-calibrated and distortion-corrected (the 
$e$ versions of the images).
Where possible, we used images which were also astrometrically
aligned (the $g$ versions of the images); however, either 
version of the images produces identical results.
Before performing photometry, we manually removed any foreground
or background objects, or instrumental artifacts such as scratches
or cosmic ray hits.  A small subset of the images also contained low-level
`stripes', and these were removed by fitting a 
Gaussian function to the profile of the `stripe' 
\citep[cf.][]{kuchinski00}.  Any residuals from 
this fitting process were found to be comparable to or weaker than variations
in the background photographic fog level, and thus they did not 
contribute significantly to the overall error budget.  We refer the reader to 
\citet{stecher97} for more details on 
the UIT data products.

Integrated fluxes for the galaxies were measured from the calibrated 
images, using the {\sc iraf}\footnote{{\sc iraf} 
	is distributed by the National Optical Astronomy Observatories,
    which are operated by the Association of Universities for Research
    in Astronomy, Inc., under cooperative agreement with the National
    Science Foundation.} task {\sc phot}.  
Magnitudes were calculated assuming:
\begin{equation}
m_{\rm UV} = -21.1 +
2.5\log_{10}(f)
\end{equation}
where $f$ is the flux in 
ergs\,s$^{-1}$\,cm$^{-2}$\,\AA$^{-1}$
\citep{stecher97}. 
The sky level was estimated
using an annulus around the galaxy, or in the case of galaxies
which were large enough to fill a significant portion of the frame, 
using boxes in areas which were free of galaxy emission.  
Our raw \fuv and \nuv magnitudes are presented in Table
\ref{tab:mag}, along with the aperture used to determine the
magnitudes.

The errors in these magnitudes are dominated by uncertainties in measuring 
the background level of the images.  These uncertainties were estimated by 
performing photometry on blank areas of the program image using the
same sized apertures as for the galaxy, or in the 
case of large galaxies, by using the dispersion in the average sky levels
measured in 20$\times$20 pixel (22.4$\times$22.4 arcsec) boxes, 
in order to measure the background variation on different
spatial scales.  The scaling factors between these dispersions and the
error in the aperture photometry were derived empirically from cases
where both were available.  These were found to be larger than
expected from Poisson statistics by approximately a factor of four,
indicating that the errors are correlated on large spatial scales
(not unexpected given the nonlinear nature of the photographic detector
used with UIT).  This algorithm reproduces the errors in the 
blank-sky aperture photometry to better than a factor of two,
over a large range of aperture area.
Other sources of error, including pixel-to-pixel noise, imperfect masking 
foreground and background objects, or cosmetic feature residuals are much
smaller than the error in the background fitting discussed above.

We devoted a considerable effort to testing for errors due to 
nonlinearity in the calibrated data.  Our tests show that any residual 
nonlinearity at high surface brightness is unimportant for the images
used in our analysis.  However, comparisons of 
long and short-exposure images of the same fields show a significant
nonlinearity a faint levels, with discrepancies of $\sim$20\% 
at exposure levels lower than 50 ADU  
\citep{stecher97}.  Our own tests confirmed this effect, but we also
found a significant variation in the magnitude of the nonlinearity 
within the data set; therefore we cannot derive an
appropriate correction for this effect.  The net effect of the nonlinearity
is to underestimate the total flux in short-exposure images by 
up to 0.2 mag when compared to long-exposure images.  We avoided
this problem by not using the short-exposure images in our analysis.
However the same effect, if present,
would also cause us to underestimate the UV fluxes in the 
long-exposure images, if a significant fraction of the emission arises
from low surface brightness regions.  Our tests
demonstrate that all but the brightest galaxies have significant
contributions from these regions, and this may lead to a systematic
underestimate of the UV fluxes by between 0.1 and 0.2 mag. 

We have compared our photometry with published \fuv data.
We constructed a weighted average of the B1 and B5 magnitudes 
(where available), and adopted this as a representative \fuv magnitude.
As a check of our internal accuracy, we have compared our
magnitudes with those derived from the same images
by \citet{waller97}.
The mean zeropoints of the two sets of magnitudes are identical, with a
median difference from galaxy to galaxy of 0.14 mag.  The RMS difference
is substantially larger (0.44 mag), but when two highly discrepant galaxies 
are excluded the RMS deviation is 0.24 mag.  We believe that the differences
largely reflect the careful background fitting in our analysis, though
the typical estimated uncertainties in our UV magnitudes are still
larger than $\pm0.1$ mag.

\begin{figure}[tb]
\epsfxsize=\linewidth
\epsfbox{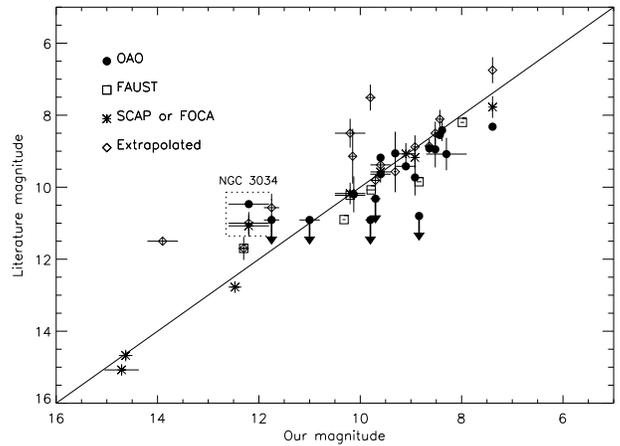}
\caption{\label{fig:comp}  
Comparison of our total \protect\fuv UIT magnitudes with total
\protect\fuv magnitudes from the literature, assuming a common 
zero point of $-$21.1.  We compare with the magnitudes from 4
UV telescopes: OAO (solid circles), FAUST (open squares), 
SCAP and FOCA (asterisks).  We also compare with the extrapolated `total' 
\protect\fuv magnitudes of \protect\citet[][open diamonds]{rifatto95}.  
NGC 3034 (M82) is indicated separately: because it is so 
highly reddened, 
its \protect\fuv flux is highly passband-dependent.  }
\end{figure}

In Figure \ref{fig:comp} we compare our magnitudes with independently
measured \fuv magnitudes
from other instruments: OAO-2 \citep[1650\,\AA;][]{code82}, FAUST
\citep[1550\,\AA;][]{deharveng94}, 
SCAP \cite[2000\,\AA;][]{donas87} and FOCA 
\citep[2000\,\AA;][]{donas90}.  We have also compared our magnitudes with 
`total' extrapolated \fuv magnitudes from \citet{rifatto95}.  In summary, 
we find reasonable agreement with the raw magnitudes from the literature, 
but with a mean offset of $-$0.23 $\pm$ 0.07 mag
(excluding NGC 3034, which is very faint in the \fuv
and has a steep UV spectrum, due to strong extinction effects, 
so the comparison for NGC 3034 is highly passband dependent).
We believe that the offset, which is significant
at the 3$\sigma$ level, is due primarily to aperture effects in
the published magnitudes, many of which were derived from apertures
smaller than or comparable to the galaxy sizes \citep{rifatto95}.
This interpretation is supported
by our comparison with the `total' extrapolated magnitudes
of \citet{rifatto95}.  We find good agreement
between most of their magnitudes and ours, with a standard
deviation of $\sim$\,0.2 mag and no offset.  We do find, however, 
that some of the largest galaxies in our sample have \fuv magnitudes 
which strongly differ (by up to 2.5 mag) from \citet{rifatto95}.  However
almost all of those cases involved very large extrapolations of 
IUE fluxes (measured in 10$\arcsec \times 20\arcsec$ 
apertures) to `total' magnitudes in the Rifatto et al. database.
This underscores the need to exercise extreme care when applying
UV data from the literature.   
In summary, we find that 
our magnitudes are accurate to $\sim$0.2$-$0.4 magnitudes, 
with no significant offset from the literature calibrations, once
aperture effects have been taken into account.

The \fuv magnitudes in Table \ref{tab:mag}
are used later to calculate SFRs.  The latter are calibrated in 
terms of luminosity per unit frequency, so 
we converted the \fuv absolute magnitudes (which are based on
$F_\lambda$) to $F_\nu$, assuming an effective wavelength for the 
B1 and B5 filters of 
$\lambda_{\rm eff,FUV} = 1567${\,\AA} (this is accurate to 
better than $\pm$6\%).  We corrected this flux for Galactic
extinction following \citet{sfd}, and
adopting the Galactic extinction curve of \citet{gordon97}.

Later, to increase the galaxy sample size, we include 33 galaxies
observed by the balloon-based FAUST telescope 
\citep{deharveng94} at 1650{\,\AA} with existing literature 
\ha fluxes.  These data have somewhat larger \fuv uncertainties,
and are subject to modest aperture effects.  These data are not
a primary focus of this paper (so they are not included in the 
tables of galaxies), but are simply included to confirm the trends
observed with the UIT galaxies with an independent sample of 
galaxies.  

\subsection{\hans, Radio Continuum, and Far-Infrared Data}

The primary goal of this paper is to compare \fuv and \ha SFRs, 
and explore the effects of dust on both of these indicators.
Consequently, we compiled measurements of the integrated fluxes of 
the galaxies in \hans, radio continuum, and \fir (as  
a probe of dust content).   The data, as discussed below, are
listed in Table \ref{tab:sfi}.  The table also lists total absolute
$V$ magnitudes and the sources of the \hans, radio, and \fir data.
The $V$ magnitudes and the \ha fluxes have been corrected for
Galactic extinction following \citet{sfd}.

The \ha fluxes come from a wide variety of sources in the literature.
In many cases more than one source was available, and in these cases
a weighted average was derived, with double weight given to fluxes
derived from CCD imaging (as opposed to photoelectric aperture 
photometry).  A few of the fluxes were measured from unpublished
CCD imaging obtained by the authors, and are published here for 
the first time.  These images have been flux-calibrated and 
were continuum-subtracted using scaled narrow-band continuum images.
Some of the data were obtained with narrowband filters which included
only the \ha line, but most were obtained with broader filters which
also included the \nii $\lambda$6548,6583 lines.  In many cases
the average $[$\ion{N}{2}$]$/\ha flux ratio has been measured from published
\hii region spectrophotometry, or from integrated spectra of the galaxies
\citep{k92}.  Otherwise, for 40\% of the galaxies, the 
\ha $+$ \nii fluxes have been corrected for contamination by
the \niil line following the prescription of \citet{k83a}.  
Total fluxes were multiplied by a factor
of 0.75 for spirals and 0.93 for types Sm and later.  Note that use of
this correction is a conservative assumption: adoption of 
a constant \nii fraction would serve to enhance the trend in \ha to
\fuv ratio which we explore later in \S \ref{sec:sfr}.

In \S \ref{sec:atten} we also use the thermal radio continuum from 
a subset of our sample of galaxies to investigate the behavior of 
the attenuation at \hans.  The integrated radio continuum flux of a 
galaxy is a typically a combination of non-thermal radio emission 
from cosmic-ray electrons and 
supernova remnants (with typically
a steep spectral index $S_\nu \propto \nu^{-0.8}$, roughly) 
and thermal bremsstrahlung emission
from electrons in ionized gas (with a much shallower spectral
index $S_\nu \propto \nu^{-0.1}$).  Unfortunately the non-thermal
component is dominant at centimeter wavelengths, typically accounting
for more than 90\% of the total flux at 1.4 GHz 
\citep{condon92,niklas97}.  Consequently the thermal fraction can
only be measured by obtaining matched-beam multi-frequency observations,
fitting the non-thermal spectrum at low frequencies and extracting
the (small) thermal component.  Reliable data are only available for
a small sample of galaxies, and for fewer yet among our sample.
For our analysis we have derived thermal radio fluxes at 1.4 GHz, 
using thermal fractions at 1 GHz derived
by \citet{niklas97} in conjunction with the
radio luminosities from the literature.  \citet{niklas97}
carefully decomposed the thermal and nonthermal
components of the radio spectra of 74 galaxies with radio fluxes 
at a minimum of 4 frequencies using a two component fit.
This sample includes 13 galaxies in common with our sample, with thermal
fractions significant at better than $1\sigma$.  
We have added a thermal fraction at 1.465 GHz
for DDO 50 from \citet{tongue95}, 
a thermal fraction at 4.9 GHz for NGC 925 from \citet{duric88}, 
and a thermal flux at 92 GHz 
for M82 taken from \citet{carlstrom91}
and discussed at length in \citet{condon92}.
Total radio continuum fluxes for our sample with thermal fractions
were available between 1.4 and 1.5 GHz: we translated these fluxes
to a 1.4 GHz flux using 
a typical spectrum for a non-thermal continuum dominated source
($S_{\nu} \propto \nu^{-0.8}$). 

\fir fluxes from the Infrared Astronomical Satellite (IRAS) have been 
compiled and translated into total $8 - 1000$ 
{\micron} fluxes using the method of \citet{fluxratio}.  
The 12 {\micron}, 25 {\micron}, 60 {\micron} and 100 {\micron} fluxes
were numerically integrated to provide an estimate of the 
$8 - 120$ {\micron} flux.  The 60 {\micron} and 100 {\micron} fluxes
were then used to define
a dust temperature for a $\beta = 1$ dust emissivity model to
extrapolate to 1000 {\micron}.  The total IR flux determined in 
this way is typically 1.9 $\pm$ 0.4 times larger than the 
{\it FIR} estimator of \citet{helou85}
which was designed to measure the bulk of the \fir flux
from a galaxy (the range in the ratio of total IR to {\it FIR} estimator
is $0.8 - 3.3$).

The \fuvns, \hans, thermal radio, and \fir fluxes were translated
into total luminosities and powers using  
the distances listed in Table \ref{tab:sample}. 
These luminosities, along with the absolute $V$ band magnitude,
are presented in Table \ref{tab:sfi}.  The typical uncertainties
in the \ha and \fir luminosities are $\pm$20\%.  The uncertainties
in the thermal radio powers are more variable and usually 
larger: the radio errors listed in Table \ref{tab:sfi} are derived
directly from the quoted errors in the thermal fractions.

\subsection{SFR Calibrations \label{sec:sfrcalib}}

Before comparing the SFRs derived from \fuv and \hans,
we must first translate the galaxy luminosities into SFRs using
the appropriate conversions.
In this paper we adopt the calibrations of \citet{k98}
for the \fuv and \ha SFRs:
\begin{equation}
{\rm SFR\,(M_{\sun}\,yr^{-1}) = 1.4 \times 10^{-28} \,
	L_{FUV}\, (ergs\,s^{-1}\,Hz^{-1})},
\end{equation}
\begin{equation}
{\rm SFR\,(M_{\sun}\,yr^{-1}) = 7.9 \times 10^{-42} \,
	L_{H\alpha}\, (ergs\,s^{-1})},
\end{equation}

For the purposes of this analysis the absolute SFR scales
are less important that adopting a consistent set of
calibrations for the two different methods.  
Both calibrations assume a \citet{sp}
initial mass function (IMF) between 0.1 M$_{\sun}$ and 100 M$_{\sun}$.  
The overall absolute SFR scale does depend on IMF (at a factor of a 
few level at most for plausible IMFs), but the important
point is that the relative comparison of \ha and \fuv SFRs is 
quite robust.  The \ha and \fuv luminosities at a given age 
depend primarily on the shape of the upper IMF over a modest range in mass 
and are independent of the shape of the lower mass end of the IMF.
For realistic ranges of upper IMF slopes the \ha to \fuv 
ratio should not vary by more than $\sim$30\% \citep[e.g.][]{glaze99}.  

The \ha line provides a virtually instantaneous measure of the
SFR, since the dominant ionizing population consists of OB
stars with lifetimes of $<$10 Myr.  However the stars contributing
to the luminosity of a galaxy at 1550{\,\AA} cover a much wider range
of ages, so the SFR calibration must explicitly assume a SFR
history over the past 100 Myr or longer.  The calibration in eq.\ ({2})
assumes continuous star formation 
over timescales in excess of 100 Myr, an approximation which is
most appropriate for a large galaxy with essentially constant
star formation when averaged over the entire disk.  However this 
conversion is reasonably robust to variations in the recent
star formation history, as long as it has been reasonably continuous
when averaged over periods of tens of Myr.  
For more discussion of the calibrations and their uncertainties, 
see \citet{k98}.

\section{COMPARISON OF UNCORRECTED STAR FORMATION RATES} \label{sec:sfr}

\begin{figure*}[tb]
\epsfxsize=\linewidth
\epsfbox{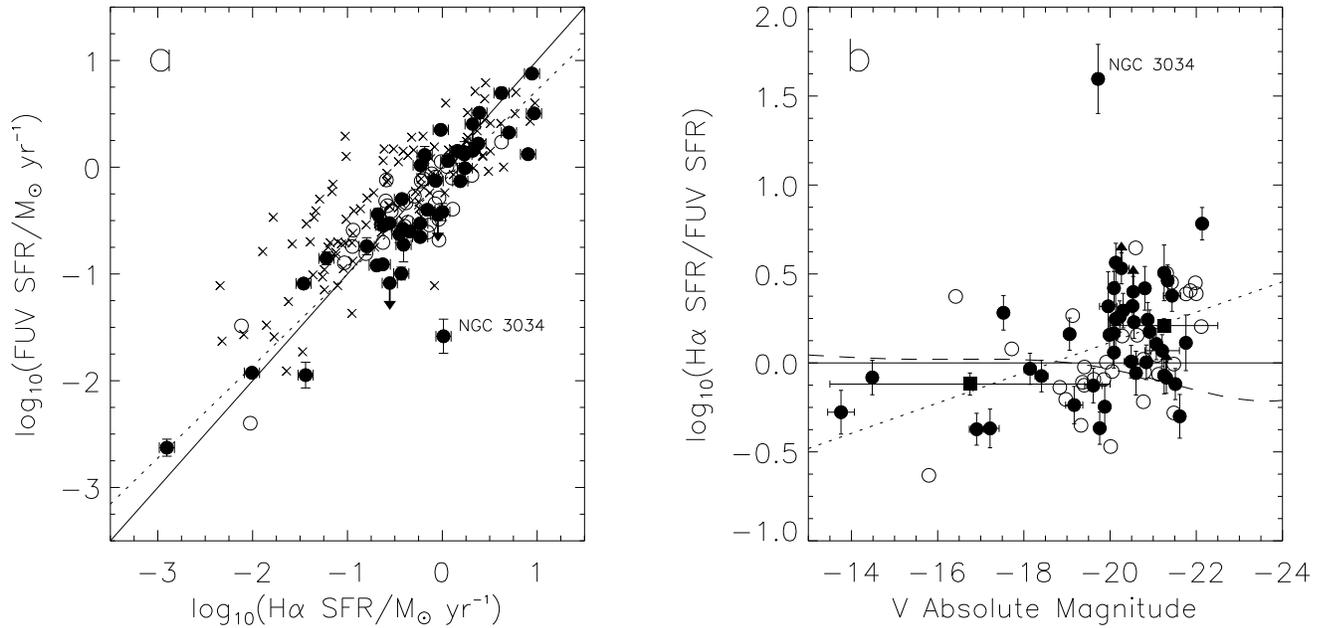}
\caption{\label{fig:sfi} Comparison of uncorrected 
\protect\fuv and \protect\ha SFRs.  The left panel shows a
direct comparison of \protect\fuv and \protect\ha SFRs
for UIT galaxies (solid circles), FAUST
galaxies (open circles) and \fuv selected galaxies from 
Sullivan et al.\ (crosses).  The 
right panel shows the \protect\ha to \protect\fuv SFR ratio as 
a function of absolute magnitude.  Arrows denote \protect\fuv upper
limits, and the solid lines denote equal SFRs, if the calibrations
in equations (2--3) are used.  Filled squares denote the mean ratios
for galaxies in two luminosity ranges, as described in the text.
The dotted line shows the best
fit to the variation of \ha to \fuv SFR ratio with 
absolute magnitude (dotted line).
The dashed line denotes the expected trend in \ha to \fuv SFR 
ratio (arbitrarily normalized to zero at $M_V = -20$) 
expected from metallicity effects alone 
\protect\citep{zkh94,sullivan00}. }
\end{figure*}

We first compare the \fuv and \ha SFRs, uncorrected for extinction, in
Figure \ref{fig:sfi}.  The left panel compares the \fuv and \ha SFRs
directly for 39 of the 50 UIT galaxies (solid circles) and
for 33 galaxies observed using FAUST \citep{deharveng94} with 
\ha data from the literature (open circles).
In addition, data for a sample of 107 \fuv selected galaxies
with $0 < z < 0.5$ from \citet{sullivan00} is shown by crosses.
These galaxies are not included explicitly in the following 
quantitative discussion, and are plotted only to allow later
comparison between our study and \citet{sullivan00}.  
The right panel displays the 
\ha to \fuv SFR ratio as a function of $V$-band absolute magnitude
for the UIT and FAUST data (solid and open circles respectively).   
A few UIT galaxies have only upper-limit \fuv fluxes, and they are 
denoted by arrows in the figure.  The solid line in both panels
corresponds to equal UV and \ha SFRs, using the calibrations in equations
(2) and (3) respectively.  
The first clear result is that uncorrected \ha and \fuv SFRs are qualitatively 
consistent with each other over three and a half orders of magnitude 
in SFR and nearly two orders of magnitude in $V$ band luminosity.  
With the exception of NGC 3034 (M82), 
the \ha and \fuv SFRs for the UIT and FAUST data are generally consistent, 
with an average offset of 0.08 $\pm$ 0.04 dex, in the sense
that the \ha SFRs are $\sim$20\% higher than the UV-derived SFRs on average.
The RMS dispersion is 0.28 dex (a factor of 1.9),
significantly higher than would be expected from the uncertainties
in the individual SFRs.    

Figure \ref{fig:sfi} also shows a significant 
trend in the ratio of \ha to \fuv SFRs with absolute magnitude.  Although
our sample is small, a Spearman rank correlation test shows that the
departure is significant
at greater than the 99.9\% level.  If we divide the sample at $M_V$ = $-$20,
the \ha to \fuv ratio for the lower luminosity galaxies is
$-$0.10 $\pm$ 0.05 dex as compared to $+$0.18 $\pm$ 0.04 dex for the more 
luminous subsample.  This corresponds to a difference (between
the lower and higher luminosity galaxies) of a factor of two.

This systematic change in the \hans/\fuv SFR ratio with SFR and galaxy
luminosity is qualitatively consistent with the previous findings of
\citet{sullivan00} 
and our results provide additional insights into
the origin of this effect.  The \fuv and \ha data analyzed by 
\citet{sullivan00} were obtained with mismatched apertures, and
the authors expressed some concern that aperture sampling effects might 
account for part of the trend in \hans/\fuv ratio.  
However, our data are largely 
free of aperture bias and we observe a qualitatively 
similar trend in \hans/\fuv SFR ratio (panel a of Fig.\ \ref{fig:sfi}),
which argues for a physical origin for the trend.  

However, our results do differ from 
those of \citet{sullivan00} in a few respects.
The uncorrected \fuv SFRs of low SFR galaxies from 
\citet{sullivan00} are larger than the raw \ha SFRs by factors of a few:
this led \citet{sullivan00} to the conclusion that the effects 
of bursts in SFR dominate their data.  
The \ha emission traces the SFR over
the past 5--10 Myr, whereas the UV flux reflects a weighted average of 
the SFR over the past $\sim$100 Myr.  If the evolution of a dwarf galaxy 
is characterized by 
brief bursts (duration $\ll$ 100 Myr) separated by relatively long quiescent
periods, then one will tend to observe most of the galaxies between
bursts, when the \ha luminosity will yield a systematically lower SFR
relative to the UV emission.  
This effect is not seen at a significant level in our sample.
It is likely that the differing behaviors of the samples 
is due to the different selection criteria.  \citet{sullivan00} selected
galaxies in the \fuv (favoring galaxies with 
increased present day SFRs).  On the other hand, 
the UIT imaged a sample of well-known, often
optically-bright galaxies (optical
selection would less weight on recent star formation).  
Because the UIT selection is relatively ill-defined, 
we do not attempt to quantitatively address the 
different selection biases between the UIT sample
and that of \citet{sullivan00}; however, it is clear
that \fuv selection would favor the inclusion of a large population 
of preferentially UV-bright starbursting galaxies.
Another possibility is that modest evolution takes place between
$0 < z < 0.5$ in the lowest SFR galaxies only \citep[e.g.][]{broad88,lilly93}.
Metallicity effects on the \ha and \fuv emission of the stars
are unlikely to account for any of the trend.  More luminous galaxies
tend to possess systematically higher metal abundances \citep{zkh94};
but, as \citet{sullivan00} discuss, the models of 
\citet{leitherer99} show that higher abundances
tend to {\it reduce} the ionizing luminosity relative to the \fuv
brightness (dashed line in Fig.\ \ref{fig:sfi}), 
contrary to the trend observed in Fig. \ref{fig:sfi}.

In low-luminosity galaxies the uncorrected UV-based SFRs are systematically 
higher than those measured from \ha by 25\% $\pm$ 12\%. 
If this were to be explained
by differential extinction effects it would require a {\it lower} 
attenuation of the UV radiation relative to \hans, a counter-intuitive
result made all the more unlikely by the generally low extinctions
observed in most dwarf galaxies.   
As mentioned earlier, 
\citet{sullivan00} suggested that the higher UV/\ha
ratios are due to temporal variations in the SFR in these low-luminosity
galaxies.  This scenario fits in well with observations of 
dwarf galaxy star formation histories, which suggest large variations
in SFR over the galaxy's lifetime \citep[e.g.][]{dohm98,tolstoy98}.
Although our sample is far too small to test the scenario directly, it
provides a plausible explanation for our observations also.
However, note that it is difficult to rule out inconsistencies
in the SFR calibration at this level 
\citep[e.g.\ \S \ref{sec:sfrcalib}][]{k98}.

In higher luminosity galaxies the trend is reversed, i.e. the uncorrected
\ha luminosities yield SFRs that are on average 50\% larger than those derived
from the uncorrected \fuv fluxes.  \citet{sullivan00} also observed
this trend, attributing it to temporal effects; in
this case postulating that the most luminous \hans-emitting galaxies
are preferentially observed at the peak of the starburst, when the
\ha luminosity is expected to yield a systematically higher SFR
\citep[also see][]{glaze99}.
However, we observe the same excess \ha emission (relative to UV) in
a sample that is dominated by normal star-forming disk galaxies, with
no evidence for current or recent starbursts.  This implies that 
temporal variations in SFRs cannot account for the inconsistency in SFRs,
at least not in our sample, and instead we tentatively attribute the
difference to higher attenuation of the \fuv emission relative to \ha
\citep[as suggested by e.g.][]{glaze99,yan99,moorcroft00}.
In \S 4 we compare the \fuv and \ha attenuations directly and confirm
this tentative conclusion.  Although temporal effects may play a 
role in accounting for the systematic difference in \ha and \fuv SFRs
observed in high-redshift galaxies, our results strongly suggest that dust
extinction effects should not be ruled out.

\section{EFFECTS OF DUST ATTENUATION} \label{sec:atten}

Dust attenuation strongly affects the 
optical and UV radiation from star-forming galaxies, as 
evidenced by the large fractional \fir luminosities of 
most spiral galaxies \citep{xu95}, by the substantial
reddenings of the Balmer lines in the brightest \hii regions
and in the integrated spectra of the galaxies themselves
\citep[e.g.][]{k92,zkh94,wang96}, and by the observed correlations
between UV spectral slope and \fuv vs \fir properties of galaxies
\citep{calzetti94,heckman98,meurer95,meurer99}.  In the previous section
we tentatively attributed the factor of 1.5 discrepancy between
uncorrected UV and \hans-derived SFRs to excess attenuation in 
the ultraviolet; in this section we derive direct constraints on
the \fuv and \ha attenuations to test this hypothesis.
We deliberately apply some of the methods that are most commonly applied
to high-redshift objects, so we can use the UIT sample to assess
the reliability of these extinction correction schemes when
independent information on the SFRs is available.

\subsection{Constraints on FUV Attenuation}

The most common methods that been employed to estimate UV attenuation
in external galaxies are to scale the attenuation 
by the mean column density of \hi
(or \hi $+$ H$_2$) gas \citep{donas84,donas87}, or to apply an empirical 
correlation between the UV spectral slope, Balmer decrement,
and total attenuation \citep[e.g.][]{calzetti94}.  More recent
work on the use of column densities to predict UV attenuation shows
a disappointing scatter \citep{buat92,buat96}, and most workers estimate
the extinction corrections using the spectral slope method.  We
have used the empirical attenuation curve determined  
for a sample of 39 starburst galaxies 
by \citet{calzetti94} to derive
approximate \fuv attenuation corrections for our sample.

The \citet{calzetti94} method effectively uses an empirically
determined extinction law to relate the attenuation at a given
wavelength to the slope of the UV spectral energy distribution.
In order to measure the spectral slopes of our galaxies we 
measured the \nuv magnitudes of a subsample of 
15 galaxies (13 of which have \fuv magnitudes also)
which were observed using Astro-1.  We then used the 
\fuv and \nuv magnitudes of these 13 galaxies to determine the slope
of their UV continuum $\beta$, following Calzetti et al.  
We have then applied their
method to determine the \fuv attenuation at
1567{\,\AA} $A_{\rm FUV}$:
\begin{equation}
A_{\rm FUV} = 0.5 k_{\lambda} (\beta+1.71)/1.88,
\end{equation}
where $k_{\lambda} = 8.66$ at 1567{\,\AA} \citep{calzetti94}.

Although this analysis allows us to characterize the approximate
\fuv extinction properties of our (sub)sample, the derived
attenuations are very sensitive to small errors in the ${\rm FUV - NUV}$
colors, and hence the values derived for individual objects are
not very meaningful.  Another concern is that the relation given 
above was determined using small-aperture observations of a sample of 
UV-bright starburst galaxies, and it may not be entirely 
appropriate for normal star-forming
galaxies over a wide range in mass and SFR.  In particular,
\citet{clump2} show 
that \fuv color-based extinction estimates
are strongly sensitive to the shape of 
the \fuv extinction curve, and that Calzetti et al.'s
\markcite{calzetti94}(1994) attenuation law necessitates
a SMC bar-type extinction curve (although 
see \citet{granato00} for a different viewpoint).
Therefore, a \fuv color-based attenuation may 
not be appropriate for spiral galaxies: we test
this in \S \ref{sec:const}.

\subsection{\ha Attenuation Measurements}

To calculate the attenuation at \ha ($A_{\rm H\alpha}$), 
we combined our \ha fluxes
with thermal radio continuum fluxes for a subset of 
16 UIT and 5 FAUST galaxies 
with reliable radio data.  The two fluxes are directly correlated,
with only a weak dependence on nebular electron temperature:
$\Delta A_{\rm H\alpha} = 1.475 \log_{10} (T_e/10^4 {\rm K})$, 
where for a reasonable range in true \hii region electron temperature 
$T_e$ from 8000K to 12000K the error in the attenuation estimate is $\mp 0.15$
mag \citep{condon92}.  We assume $T_e = 10000$K.
Following \citet{condon92}, we translate the
thermal radio flux into an \ha luminosity via:
\begin{equation}
L_{\rm H\alpha, predicted} ({\rm ergs\,s^{-1}}) = 
8.3 \times 10^{13} L_{\rm thermal} ({\rm ergs\,s^{-1}\,Hz^{-1}})
\end{equation}
where  $L_{\rm H\alpha, predicted}$ is the predicted \ha luminosity from
a given thermal radio continuum luminosity at 1.4 GHz $L_{\rm thermal}$.

As is the case with the \fuv attenuations derived from the $\beta$
method, the \ha attenuations derived in this way are subject to
large uncertainties for individual galaxies, in this case due to
the errors associated with separating the non-thermal and thermal
components to the radio luminosity (see \S 2.2).  

One common method for estimating the attenuation at \ha is
to compare the ratio of the \ha and \hb fluxes with the 
theoretical value of 2.86 \citep[at $T_e = $10000 K;][]{cd86}.  This gives the
reddening between \ha and H$\beta$: with the assumption 
of an attenuation curve it is possible to then estimate
the attenuation at \hans, or indeed, any other wavelength. 
This approach has known limitations.
Generally, Balmer decrement measurements are only available for
bright \hii regions within the galaxies, and applying this mean 
extinction to the galaxy as a whole is a rough approximation at best.
Moreover, the radio-derived attenuation 
of \hii regions in the LMC are larger than those expected from 
the Balmer decrements using a naive screen 
model, indicating a grayer extinction curve \citep[probably due to the 
effects of geometry; e.g.][]{cd86}.  This method is commonly used despite
its limitations as both \ha and \hb are easily accessible, 
observationally speaking, and often can even be measured off of the 
same spectrum (thus minimizing e.g.\ aperture mismatches or calibration 
uncertainties).

We have constructed \hahb ratios for 16 UIT and FAUST galaxies
with radio-based \ha attenuations.  These were derived either from
spectra of either the entire galaxy \citep[e.g.][]{k92} or 
are the average of published \hahb for individual \hii regions.
These \hahb values are converted into an attenuation estimate 
assuming an intrinsic \hahb of 2.86, and assuming a galactic dust
screen model.  We find that the \ha attenuations derived from 
radio data are somewhat larger on average than the \ha attenuations 
derived from Balmer decrements (0.2$\pm$0.2 mag), with a scatter of nearly
0.8 mag.  This modest offset is consistent with a grayer extinction 
curve, as found by \citet{cd86} for \hii regions in the LMC.
Given the modest numbers of galaxies with Balmer decrements, 
and bearing in mind the systematic uncertainties inherent
to Balmer decrement-based attenuations, we do not consider 
these further, except to note that these attenuations are 
consistent with the radio-based values, with much scatter.

\subsection{Comparison of FUV and \ha Attenuation Distributions}

\begin{figure*}[tb]
\epsfxsize=\linewidth
\epsfbox{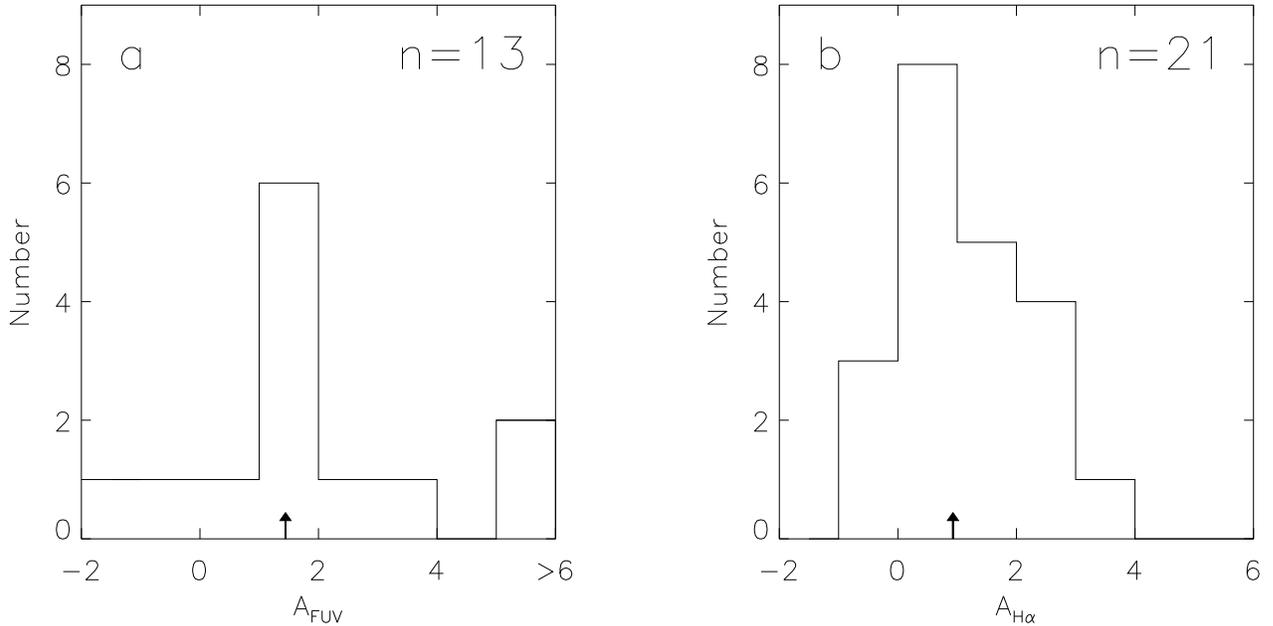}
\caption{\label{fig:hist}  Distributions of the 
\fuv (panel a) and \ha (panel b) attenuations for two different
subsamples of 13 and 21 galaxies respectively.  Arrows denote the median 
attenuation. }
\end{figure*}

We compare the \fuv and \ha attenuation distributions in 
Fig.\ \ref{fig:hist} for the respective subsamples of 
13 and 21 galaxies.  Arrows denote the median attenuation.
The two
galaxies with derived $A_{FUV} > 6$ mag (NGC~2551, NGC~3034)
have spectral slopes $\beta > 0$, and lie 
outside the calibration range of \citet{calzetti94}.

Figure \ref{fig:hist} shows that the 
distributions of \ha and 1567{\,\AA} \fuv attenuations are 
qualitatively similar.  The median \fuv attenuation is   
1.4 mag as compared to 0.9 mag for \hans.
The somewhat higher \fuv attenuations are consistent
with the systematic offset in uncorrected SFRs discussed earlier,
although the formal uncertainties in the median values are roughly
$\pm$0.3 mag in both cases, so the difference 
between \ha and \fuv values is only marginally
significant.  Furthermore, these samples only have two galaxies in common 
(NGC 3031 and NGC 3034), so it is dangerous to draw any firm conclusions
from the relative attenuation distributions.

\subsection{Comparison with FIR Properties} \label{sec:const}

\begin{figure*}[tb]
\epsfxsize=\linewidth
\epsfbox{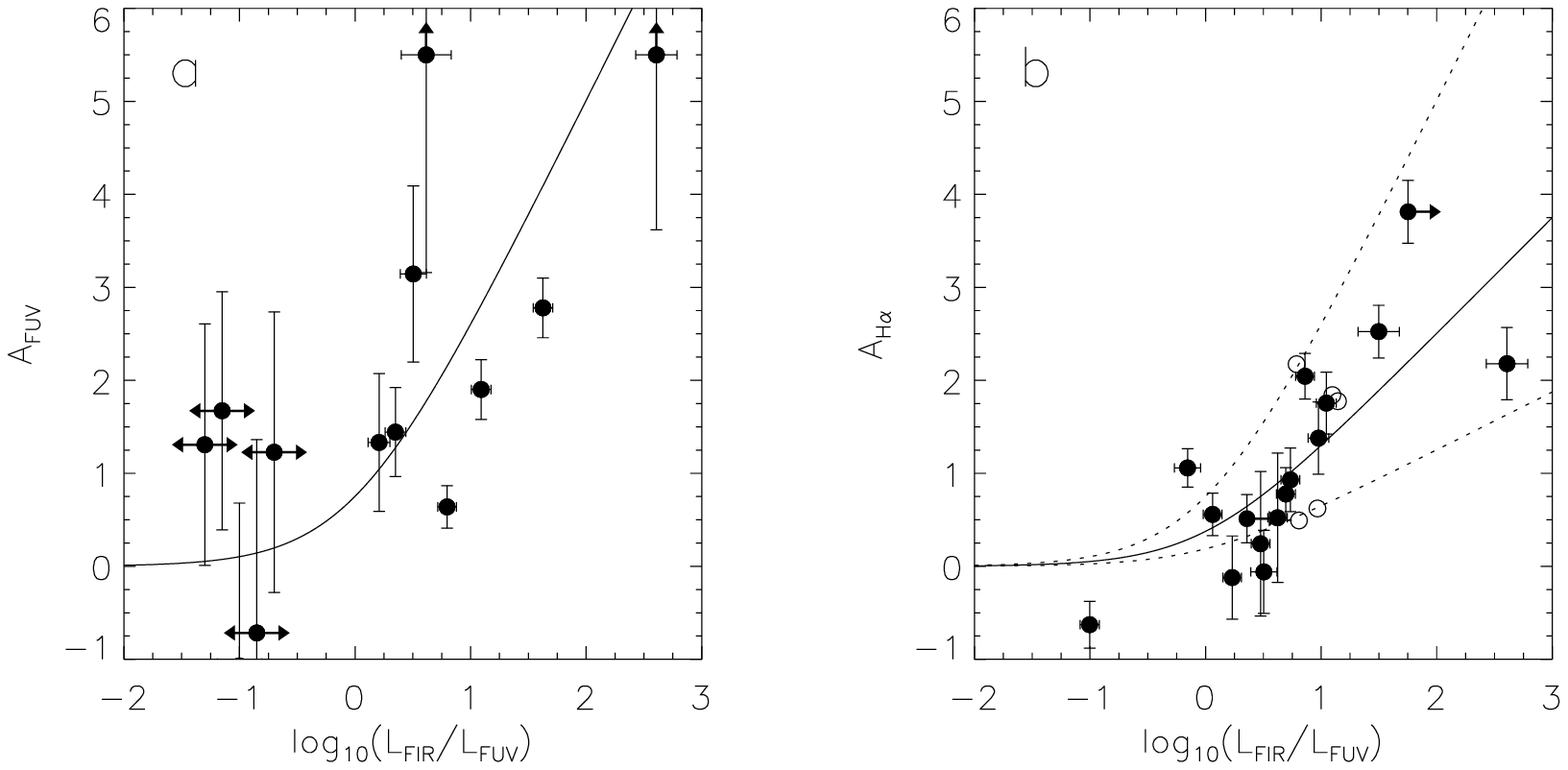}
\caption{\label{fig:atten} Comparison of the \fuv (panel a) 
and \ha (panel b) attenuation 
with \firfuvns.
Arrows in panel a denote galaxies without 
\fir luminosities; however, these galaxies are likely on the whole to have low 
\firfuv (and have $A_{\rm FUV}$ values which are consistent with zero in
any case).   Arrows in panel b denote \fuv
luminosity upper limits.
The solid line shows the expected relationship if the \fir 
luminosity was produced solely by absorption of \fuv radiation, and
produced an attenuation at \ha half as large as
that at 1567{\,\AA}.  Dotted lines denote the relationship 
if the \ha attenuation was as large (upper line) or only 
a quarter as large (lower line) as the \fuv attenuation.}
\end{figure*}

One interesting question is whether or not there is evidence
against a single, representative attenuation
for our sample of star-forming galaxies.  In other words, 
is the spread in the attenuations in Fig.\ \ref{fig:hist}
just a random scatter, or does the spread indicate
a systematic variation in spiral galaxy attenuation?
There is already ample evidence for an increase in UV
attenuation with increasing \fir luminosity 
\citep[e.g.][]{wang96,meurer99,buat99}.  However, 
there has been no evidence for an increase in \ha attenuation 
as a function of \fir luminosity: there is 
evidence for increased \ha reddening \citep{calzetti94,wang96} or a decreased
\ha to \fir luminosity ratio with \fir luminosity \citep{cram98}, but no direct
evidence for a change in \ha attenuation.

To test for systematically varying attenuation in these galaxies, 
Fig.\ \ref{fig:atten} plots the color-based \fuv and 
\ha attenuations derived above
against a different indicator of UV attenuation:
the \fir to \fuv luminosity ratio.  This ratio 
is subject to its own set of systematic
uncertainties (e.g.\ the possible influence of old stellar population 
heating of the dust, the assumption of a flat \fuv spectrum in
terms of flux per unit frequency,
or dust heating from ionizing \fuv radiation), however
the \firfuv should suffice to 
rank galaxies roughly by the amount of UV attenuation, 
at the very least.

In Fig.\ \ref{fig:atten}, we show the 1567{\,\AA} \fuv (panel a)
and \ha (panel b) attenuations against
\firfuvns.  The solid line denotes
the expected relationship if the \fir 
luminosity was produced solely by absorption of \fuv radiation, and
if the attenuation at \ha was half as large as that at 1567{\,\AA}.
Dotted lines denote the relationship 
if the \ha attenuation was as large (upper line) or only 
a quarter as large (lower line) as the \fuv attenuation.
In panel a, we can see that the evidence 
for a correlation between the UV spectral slope-based attenuation
and \firfuv is inconclusive.  
This may indicate that while $\beta$ for spiral galaxies may give 
a rough measure of the \fuv attenuation, it may not be accurate
on a case-by-case basis (perhaps due to variation in 
the shape of the \fuv attenuation curves).  
However, only six of the thirteen galaxies have both accurate
$A_{\rm FUV}$ and \firfuv values: five galaxies lack significant 
\fir detections 
(we plot these galaxies with $\log_{10}({\rm FIR/FUV})\,\sim\,-1$
as they are likely to have reasonably low \firfuvns), and the other
two galaxies have UV spectral slopes outside the calibration 
range of \citet{calzetti94}.
Clearly more \nuv and \fuv magnitudes of a larger
sample of galaxies are required to properly address this question
with any degree of certainty.

In contrast, the \ha attenuations (panel b; determined from the 
\ha to thermal radio continuum ratio, for galaxies where
the thermal radio was detected at better than 1$\sigma$) appear to
correlate more strongly with \firfuvns.  Somewhat 
surprisingly, the solid line, which denotes the 
relationship between attenuation and \firfuv
(assuming that the \ha attenuation is half of that in the \fuvns, 
and assuming that \firfuv is essentially a \fuv attenuation indicator
for this sample) describes the data rather well.
This should be taken with 
some caution, as there is likely a contribution to the dust heating from 
older stellar populations (or absorption of the ionizing
continuum before it can ionize hydrogen in \hii regions), which
would add scatter and/or a systematic bias to this comparison.

Despite the modest sample size, the correlation between 
$A_{\rm H\alpha}$ and \firfuv dispels the notion of 
a constant \ha attenuation: there is a large range of
attenuations in star-forming galaxies, and the \ha and
UV attenuation in a given galaxy are at least loosely correlated.
Furthermore, the range in \ha attenuations is at least 4 mags in our
sample of galaxies, indicating that emission line-based SFR estimates
of very highly obscured galaxies
may be up to 1 or 2 orders of magnitude lower than the true SFRs. 
The UV is affected by more attenuation, and the problem may be even more
severe for the UV-based SFRs of highly obscured galaxies.  
This scenario is qualitatively
consistent with that outlined in by e.g.\ 
\citet{buat99}, \citet{wang96}, or \citet{heckman98}
where more massive, higher
SFR galaxies have more attenuation.
This would explain the factor of 1.5
offset between \ha and \fuv SFRs for high luminosity galaxies (in this
picture, low luminosity galaxies have comparable 
SFRs in the \ha and \fuv because the effects of dust
are small for both SFR indicators), 
and the curve in panel b of Fig.\ \ref{fig:atten}.
Furthermore, this scenario is roughly consistent with the 
distributions of \fuv and \ha attenuations in Fig.\ \ref{fig:hist}.

\section{CONCLUSIONS} \label{sec:conc}

We have performed photometry on UIT \nuv and \fuv images of a sample
of 50 star-forming galaxies.  Comparison with literature \fuv magnitudes
suggests that our magnitudes are accurate to better than 0.4 mag.  
Combining these \nuv and \fuv
magnitudes with literature \hans, \firns, and
thermal radio continuum measurements, we have found the following.
\begin{itemize}
\item Raw \fuv and \ha SFR estimates are consistent to within a factor
	of two.  At low luminosities, \ha SFR estimates are around
	25\% lower then \fuv SFR estimates, indicating 
	either inconsistencies in the SFR calibrations, 
	or the effects of low-level
	bursts of star formation.  At higher luminosities, 
	\ha SFR estimates are $\sim\,$1.5 times higher than 
	\fuv SFR estimates, with a factor-of-two scatter.  
	This indicates that \fuv attenuation is, on average, larger
	than \ha attenuation.  Furthermore, the difference in 
	\ha to \fuv SFR ratio between high and low luminosity 
	galaxies suggests that there are differences in 
	extinction properties which correlate at least
	loosely with luminosity.
\item For subsets of our sample of galaxies, 
	we have constructed \fuv attenuation estimates using 
	the UV spectral slope as an attenuation indicator
	\citep{calzetti94}, and \ha attenuation estimates
	by comparing \ha and thermal radio continuum luminosities.
	The median \fuv attenuation is somewhat larger
	than the median \ha attenuation, although 
	with marginal significance.
\item Comparison of the \ha attenuation with the \firfuv 
	for our galaxies demonstrates 
	that there is a broad range in \ha attenuation.  If \firfuv is 
	driven mainly by attenuation, there
	are suggestions that the \ha attenuation correlates
	broadly with the \fuv attenuation, 
	and is around a factor of two lower than the \fuv attenuation. 
\item A characteristic \ha and \fuv attenuation for our sample of galaxies is 
	$\sim$ 0.9 mag and $\sim$ 1.4 mag respectively, although 
	there is a large range in attenuations in our sample, 
	ranging from 0 to 4 magnitudes at \hans. 
\end{itemize}

\acknowledgements

We thank Karl Gordon for providing the total IR estimation IDL routine, 
and for numerous helpful discussions.  We acknowledge useful 
suggestions from and discussions with Mark Sullivan, Richard Ellis, 
Gerhardt Meurer, and especially the anonymous referee.
This work was supported by NASA grant NAG5-8426 and NSF grant 
AST-9900789.
Some of the data presented in this paper was obtained from the 
Multimission Archive at the Space Telescope Science Institute (MAST).
STScI is operated by the Association of Universities for Research
in Astronomy, Inc., under NASA contract NAS5-26555.  Support for MAST
for non-HST data is provided by the NASA Office of Space Science
via grant NAG5-7584 and by other grants and contracts.
This research has made extensive use of the NASA/IPAC Extragalactic Database 
(NED) which is operated by the Jet Propulsion Laboratory, 
California Institute of Technology,
under contract with the National Aeronautics and Space Administration.

\onecolumn
\clearpage
\begin{table}
\begin{scriptsize}
\begin{center}
\caption{Sample properties {\label{tab:sample}}} 
\begin{tabular}{llcclccc}
\tableline
\tableline
{Galaxy} & {Alternative} & {RA (2000)} &
{Dec (2000)} & {RC3} & {D} & {Dist} & {Galactic}   \\
{Name} & {Name} & {hh\,mm\,ss} &
{dd\,mm\,ss} & {Type} & {(Mpc)} & {Ref} & {A$_V$} \\
\tableline
NGC 253 & \nodata	& 00 47 33.1	& $-$25 17 18	& SAB(s)c; \hii &
2.6	& 8	& 0.06  \\
NGC 628	& M74	& 01 36 41.7	& 15 46 59	& SA(s)c	&
10.0	& 1	& 0.23	\\
NGC 891	& \nodata	& 02 22 33.4	& 42 20 57	& SA(s)b; pec	&
9.9	& 5	& 0.22	\\
NGC 925 & \nodata 	& 02 27 17.0	& 33 34 43	& SAB(s)d	&
8.9	& 1	& 0.25  \\
NGC 1068 & M77	& 02 42 40.7	& $-$00 00 48	& (R)SA(rs)b; Sy &
15.1	& 1	& 0.11	\\
NGC 1097 & \nodata 	& 02 46 19.1	& $-$30 16 28	& (R)SB(r)b; Sy &
15.9	& 1	& 0.09	\\
NGC 1291 & NGC 1269 & 03 17 18.3 & $-$41 06 28 	& (R)SB(l)0/a	&
8.6	& 7	& 0.04	\\
NGC 1313 & \nodata 	& 03 18 15.4 	& $-$66 29 51	& SB(s)d; \hii &
3.9	& 1	& 0.37	\\ 
NGC 1317 & \nodata 	& 03 22 44.4	& $-$37 06 13	& (R)SAB(rl)0/a &
20	& 11	& 0.07	\\
NGC 1365 & \nodata 	& 03 33 36.4	& $-$36 08 25	& (R)SBb(s)b; Sy &
20.5	& 1	& 0.07	\\
NGC 1512 & \nodata 	& 04 03 54.3	& $-$43 20 57	& SB(r)ab	&
9.8	& 1	& 0.04  \\
NGC 1566 & \nodata 	& 04 20 00.6 	& $-$54 56 17	& (R)SAB(rs)bc; Sy &
17.6	& 1	& 0.03	\\
NGC 1672 & \nodata	& 04 45 42.1 	& $-$59 14 57	& (R)SB(r)bc; Sy &
15.4	& 1	& 0.08  \\
NGC 2146 & \nodata	& 06 18 39.7	& 78 21 23 	& SB(s)ab; pec  &
14.5	& 6	& 0.32	\\
NGC 2403 & \nodata 	& 07 36 51.4	& 65 36 09	& SAB(s)cd; \hii &
3.0	& 1	& 0.13  \\
NGC 2551 & \nodata 	& 08 24 50.2	& 73 24 44	& SA(s)0/a	&
31	& 3	& 0.09	\\
NGC 2841 & \nodata 	& 09 22 02.7	& 50 58 36	& SA(r)b; LINER/Sy &
9.0	& 1	& 0.05	\\
NGC 2903 & \nodata 	& 09 32 10.1	& 21 30 04	& SB(s)d; \hii &
6.3	& 1	& 0.10 	\\
NGC 2993 & \nodata 	& 09 45 43.8	& $-$14 22 05	& Sa pec	&
32	& 3	& 0.20	\\
NGC 3031 & M81	& 09 55 33.2	& 69 03 55	& SA(s)ab; LINER/Sy &
3.3	& 1	& 0.27  \\
NGC 3034 & M82	& 09 55 52.2	& 69 40 47	& I0; \hii	&
3.3	& 1	& 0.53  \\
NGC 3310 &  \nodata	& 10 38 45.9	& 53 30 12	& SAB(r)bc pec; \hii &
13.9	& 1	& 0.07	\\
NGC 3351 & M95	& 10 43 58.0	& 11 42 14 	& SB(r)B; \hii	&
9.0	& 1	& 0.09  \\
NGC 3389 & \nodata	& 10 48 27.9	& 12 32 00	& SA(s)c	&
24	& 10	& 0.09	\\
NGC 4038/9 & The Antennae & 12 01 55.2 & $-$18 52 44 & Pec/int &
19.8	& 1	& 0.15  \\
NGC 4156 & \nodata	& 12 10 49.6	& 39 28 22	& SB(rs)b; LINER &
90	& 3	& 0.09	\\
NGC 4214 & \nodata	& 12 15 38.7 	& 36 19 42	& IAB(s)m; \hii &
4.2	& 1	& 0.07  \\
NGC 4258 & M106 & 12 18 57.5	& 47 18 14	& SAB(s)bc; LINER/Sy &
6.8	& 1	& 0.05	\\
NGC 4321 & M100 & 12 22 54.9	& 15 49 21	& SAB(s)bc; LINER/\hii &
16.0	& 1	& 0.09	\\
NGC 4449 & \nodata	& 12 28 12.0	& 44 05 41	& IBm; \hii	&
3.4	& 1	& 0.06 	\\
NGC 4631 & \nodata 	& 12 42 05.9	& 32 32 22	& SB(s)d	&
8.4	& 1	& 0.06	\\
NGC 4647 & \nodata 	& 12 43 32.3	& 11 34 55	& SAB(rs)c	&
20	& 11	& 0.09 	\\
NGC 4736 & M94  & 12 50 53.1	& 41 07 14	& (R)SA(r)ab; LINER &
4.8	& 1	& 0.06	\\
NGC 5055 & M63	& 13 15 49.3	& 42 01 49	& SA(rs)bc; H\,{\sc ii}/LINER &
7.6	& 1	& 0.06  \\
NGC 5194 & M51	& 13 29 52.3	& 47 11 54	& SA(s)bc pec; H\,{\sc ii}/Sy &
7.3	& 1	& 0.12  \\
NGC 5236 & M83	& 13 37 00.8	& $-$29 51 59	& SAB(s)c; \hii &
3.7	& 1	& 0.22	\\
NGC 5253 & \nodata 	& 13 39 55.9	& $-$31 38 24	& Im pec; \hii &
3.6	& 1	& 0.19	\\
NGC 5457 & M101 & 14 03 12.5	& 54 20 55	& SAB(rs)cd	&
7.4	& 1	& 0.03	\\
NGC 6090 & Mrk 496 & 16 11 40.3	& 52 27 26	& Pec/int	&
117	& 3	& 0.07	\\
NGC 6946 & \nodata 	& 20 34 52.3 	& 60 09 14	& SAB(rs)cd; \hii &
6.2	& 9	& 1.13  \\
DDO 50 & Holmberg II & 08 19 06.0 & 70 42 51	& Im	&
3.1	& 2 	& 0.11  \\
DDO 66 & Holmberg IX & 09 57 32.4 & 69 02 35	& Im	&
3.3	& 1	& 0.26	\\
DDO 75 & Sextans A	& 10 11 00.8	& $-$04 41 34	& IBm &
1.45	& 4	& 0.15	\\
DDO 81 & IC 2574 & 10 28 21.2	& 68 24 43	& SAB(s)m	&
3.1	& 2	& 0.12	\\
UGC 6697 & \nodata 	& 11 43 49.2	& 19 58 05	& Im	&
90	& 3	& 0.07	\\
UGC 7188 & \nodata	& 12 11 16.4	& 39 24 08	& SBm	&
13	& 3	& 0.08  \\
CGCG 097-093 & \nodata	& 11 44 01.9	& 19 47 04 	& Irr 	&
66	& 3	& 0.07	\\ 
CGCG 097-114 & \nodata 	& 11 44 47.7	& 19 46	27	& Irr	&
111	& 3	& 0.08	\\
Mrk 66 & \nodata	& 13 25 53.7	& 57 15 05 	& BCG	&
87	& 3	& 0.04	\\
\scriptsize{IRAS 08339+6517} & \nodata  & 08 38 23.4 	& 65 07 14	& Pec; \hii	&
76	& 3	& 0.31 	\\
\tableline \\
\end{tabular}
\end{center}
\tablerefs{(1) \cite{waller97} 
	(2) \cite{sandage74} (3) Hubble flow assuming
	H$_0 = 75$ kms$^{-1}$\,Mpc$^{-1}$ (4) \cite{sakai96} 
	(5) \cite{ciardullo91} (6) \cite{hutchings90} 
	(7) \cite{crocker96} (8) \cite{puche88} (9) \cite{kara97}
	(10) \cite{shanks97} (11) Distance to the Virgo and Fornax Clusters, 
	following \cite{shanks97} }
\end{scriptsize}
\end{table}

\clearpage
\begin{deluxetable}{llll}
\tablewidth{12cm}
\tablecaption{UV exposure times {\label{tab:exp}}} 
\tablehead{
\colhead{Galaxy} & \multicolumn{2}{c}{Far-UV} & \colhead{Near-UV}  \\
\colhead{Name} &  \colhead{B1\,(s)} & \colhead{B5\,(s)} & \colhead{A1\,(s)}
}
\startdata
NGC 253 & \nodata	& 275.5		& \nodata	\\
NGC 628	& 514.5		& \nodata	& 530.5	\\
NGC 891 & 437.5,380.5   & \nodata	& 437.5	\\
NGC 925 & \nodata	& 1590.5 	& \nodata	\\
NGC 1068 & 752.5,563.5  & 987.5,752.5   & 563,129 \\
NGC 1097 & \nodata	& 1120.5,605.5 & \nodata	\\
NGC 1291 & \nodata	& 1590.5 	& \nodata	\\
NGC 1313 & \nodata	& 1070.5	& \nodata	\\
NGC 1317 & 545.5 	& 545.5 	& 545.5 \\
NGC 1365 & \nodata	& 973.5		& \nodata	\\
NGC 1512 & \nodata	& 948.5		& \nodata	\\
NGC 1566 & \nodata	& 1390.5	& \nodata	\\
NGC 1672 & \nodata	& 926.5		& \nodata	\\
NGC 2146 & 442.5 	& 443.5 	& 442.5 \\
NGC 2403 & 771.5	& \nodata	& \nodata	\\
NGC 2551 & 1060.5 	& 451.5 	& 454.5 \\
NGC 2841 & 1020.5	& \nodata 	& \nodata	\\
NGC 2903 & 548.5,348.5  & \nodata 	& \nodata	\\
NGC 2993 & 256.5 	& 541.5 	& 256.5 \\
NGC 3031 & 640.5	& \nodata 	& 639.5	\\
NGC 3034 & 270.5	& \nodata 	& 270.5	\\
NGC 3310 & 1130.5	& \nodata 	& \nodata	\\
NGC 3351 & 880.7	& \nodata 	& \nodata	\\
NGC 3389 & 1300.5	& \nodata 	& \nodata	\\
NGC 4038/9 & 880.5	& \nodata 	& \nodata	\\
NGC 4156 & 832.5,804.5 	& 1199.5 	& 239.5 \\
NGC 4214 & 1060.5	& \nodata 	& \nodata	\\
NGC 4258 & 1310.5	& \nodata 	& \nodata	\\
NGC 4321 & \nodata 	& 226.5		& \nodata	\\
NGC 4449 & 986.7 	& 493.5 	& \nodata	\\
NGC 4631 & 1140.5 	& 1560.5 	& \nodata \\
NGC 4647 & 1300.5	& \nodata 	& \nodata	\\
NGC 4736 & 1040.5	& \nodata 	& \nodata	\\
NGC 5055 & 1140.5	& \nodata 	& \nodata	\\
NGC 5194 & 1100.5	& \nodata 	& \nodata	\\
NGC 5236 & 818.5 	& 1430.5 	& \nodata \\
NGC 5253 & 725.7 	& 726.5 	& \nodata	\\
NGC 5457 & 1310.5	& \nodata 	& \nodata	\\
NGC 6090 & 966.5,842.5 	& \nodata 	& \nodata	\\
NGC 6946 & \nodata 	& 546.5		& \nodata	\\
DDO 50 & 1310.5		& \nodata 	& \nodata	\\
DDO 66 & 640.5		& \nodata 	& 639.5	\\
DDO 75 & 689.5		& \nodata 	& \nodata	\\
DDO 81 & 623.5		& \nodata 	& \nodata	\\
UGC 6697 & 753.5 	& 529.5 	& 105.5 \\
UGC 7188 & 832.5,804.5  & 1199.5 	& 239.5 \\
CGCG 097-093 & 753.5 	& 529.5 	& 105.5 \\
CGCG 097-114 & 753.5 	& 529.5 	& 105.5 \\
Mrk 66 & 775.5		& \nodata	& \nodata	\\
IRAS 08339+6517 & 1186.5 & \nodata 	& \nodata	\\
\enddata
\end{deluxetable}

\clearpage
\begin{deluxetable}{lllll}
\tablewidth{12cm}
\tablecaption{Ultraviolet magnitudes {\label{tab:mag}}} 
\tablehead{
\colhead{Galaxy} & \colhead{B1} & \colhead{B5}  & \colhead{A1} & \colhead{Aperture} \\
\colhead{Name} & \colhead{Magnitude} & \colhead{Magnitude} 
 & \colhead{Magnitude} & \colhead{(arcsec)}}
\startdata
NGC 253  & \nodata		& 8.3 $\pm$ 0.4 & \nodata	& 700 \\
NGC 628	 & 9.7 $\pm$ 0.1	& \nodata	& 10.18 $\pm$ 0.03 & 300 \\
NGC 891  & $>$ 12.5 (3$\sigma$) & \nodata	& 13.4 $\pm$ 0.4 & 100 \\
NGC 925  & \nodata		& 9.6 $\pm$ 0.2 & \nodata	& 400 \\
NGC 1068 & 10.21 $\pm$ 0.05	& 9.8 $\pm$ 0.1 & 10.35 $\pm$ 0.03 & 200 \\
NGC 1097 & \nodata		& 10.2 $\pm$ 0.2 & \nodata	& 250 \\
NGC 1291 & \nodata		& 13.9 $\pm$ 0.3 & \nodata	& 70 \\
NGC 1313 & \nodata		& 8.9 $\pm$ 0.10 & \nodata	& 300 \\
NGC 1317 & 13.25 $\pm$ 0.05	& 13.10 $\pm$ 0.15 & 13.66 $\pm$ 0.02 & 40 \\
NGC 1365 & \nodata		& 9.65 $\pm$ 0.20 & \nodata	& 300 \\
NGC 1512 & \nodata		& 11.60 $\pm$ 0.10 & \nodata	& 100 \\
NGC 1566 & \nodata		& 8.84 $\pm$ 0.05 & \nodata	& 300 \\
NGC 1672 & \nodata		& 10.35 $\pm$ 0.10 & \nodata	& 200 \\
NGC 2146 & $>$ 12 (3$\sigma$)	& $>$ 11 (3$\sigma$) & 12.9 $\pm$ 0.2 & 200 \\
NGC 2403 & 8.43 $\pm$ 0.06	& \nodata	& \nodata	& 600 \\
NGC 2551 & 14.8 $\pm$ 0.5	& $>$ 13.5 (3$\sigma$) & 14.05 $\pm$ 0.10 &70\\
NGC 2841 & 11.0 $\pm$ 0.2	& \nodata	& \nodata	& 250 \\
NGC 2903 & 10.15 $\pm$ 0.08	& \nodata	& \nodata	& 150 \\
NGC 2993 & 11.90 $\pm$ 0.03	& 11.95 $\pm$ 0.08 & 12.57 $\pm$ 0.03 & 50 \\
NGC 3031 & 9.1 $\pm$ 0.2	& \nodata	& 9.20 $\pm$ 0.05 & 700 \\
NGC 3034\tablenotemark{a} & 12.2 $\pm$ 0.4 & \nodata & 10.97 $\pm$ 0.09	&150 \\
NGC 3310 & 9.60 $\pm$ 0.03	& \nodata	& \nodata	& 150 \\
NGC 3351 & 11.75 $\pm$ 0.15	& \nodata	& \nodata	& 130 \\
NGC 3389 & 11.70 $\pm$ 0.02	& \nodata	& \nodata	& 100 \\
NGC 4038/9 & 10.32 $\pm$ 0.03	& \nodata	& \nodata	& 100 \\
NGC 4156 & 14.3 $\pm$ 0.2	& 14.5 $\pm$ 0.3 & 14.83 $\pm$ 0.15 & 34 \\
NGC 4214 & 8.92 $\pm$ 0.04	& \nodata 	& \nodata	& 300 \\
NGC 4258 & 9.6 $\pm$ 0.2	& \nodata	& \nodata	& 500 \\
NGC 4321 & \nodata		& 10.2 $\pm$ 0.3 & \nodata	& 240 \\
NGC 4449 & 8.37 $\pm$ 0.02	& 8.45 $\pm$ 0.04 & \nodata	& 200 \\
NGC 4631 & 8.66 $\pm$ 0.05	& 8.55 $\pm$ 0.10 & \nodata 	& 450 \\
NGC 4647 & 12.3 $\pm$ 0.1	& \nodata	& \nodata	& 100 \\
NGC 4736 & 9.31 $\pm$ 0.03	& \nodata	& \nodata	& 200 \\
NGC 5055 & 9.8 $\pm$ 0.1	& \nodata	& \nodata	& 250 \\
NGC 5194 & 8.52 $\pm$ 0.04	& \nodata	& \nodata	& 300 \\
NGC 5236 & 8.01 $\pm$ 0.06	& 7.98 $\pm$ 0.03 & \nodata	& 350 \\
NGC 5253 & 9.80 $\pm$ 0.10	& 9.77 $\pm$ 0.10 & \nodata	& 200 \\
NGC 5457 & 7.39 $\pm$ 0.02	& \nodata	& \nodata	& 800 \\
NGC 6090 & 13.50 $\pm$ 0.03	& \nodata	& \nodata	& 25 \\
NGC 6946 & \nodata		& $>$ 10.2 (3$\sigma$) & \nodata & 250 \\
DDO 50	& 9.71 $\pm$ 0.05	& \nodata	& \nodata	& 250 \\
DDO 66  & 14.1 $\pm$ 0.2	& \nodata	& 14.6 $\pm$ 0.2 & 75 \\
DDO 75	& 10.25 $\pm$ 0.10	& \nodata	& \nodata	& 175 \\
DDO 81	& 9.15 $\pm$ 0.15	& \nodata	& \nodata	& 350 \\
UGC 6697 & 12.44 $\pm$ 0.10	& 12.6 $\pm$ 0.2 & 13.02 $\pm$ 0.10 & 70 \\
UGC 7188 & 14.3 $\pm$ 0.2	& 14.4 $\pm$ 0.3 & 15.4 $\pm$ 0.3 & 40 \\
CGCG 097-093 & 14.62 $\pm$ 0.10	& 14.7 $\pm$ 0.3 & 15.2 $\pm$ 0.3 & 35 \\
CGCG 097-114 & 14.9 $\pm$ 0.4	& 14.6 $\pm$ 0.3 & 15.7 $\pm$ 0.3 & 30 \\
Mrk 66  & 13.18 $\pm$ 0.06	& \nodata	& \nodata	& 40 \\
IRAS 08339+6517 & 12.25 $\pm$ 0.05 & \nodata	& \nodata	& 100 \\
\enddata
\tablenotetext{a}{NGC 3034 is very faint in the B1 image: only a faint
	`plume' of far-UV light scattered off of dust is detected 
	\citep{marcum00}.}
\end{deluxetable}

\clearpage
\begin{table}
\begin{scriptsize}
\begin{center}
\caption{Luminosities: Star Formation Indicators {\label{tab:sfi}}} 
\begin{tabular}{lcccccc}
\tableline
\tableline
{Galaxy} & {M$_V$} & {UV} 
& {H$\alpha$}  & {FIR} & {Radio}  & {Refs} \\
{Name} & {}
& {($10^{27}$ ergs\,s$^{-1}$\,Hz$^{-1}$)}
& {($10^{41}$ ergs\,s$^{-1}$)}  
& {($10^{43}$ ergs\,s$^{-1}$)} 
& {($10^{27}$ ergs\,s$^{-1}$\,Hz$^{-1}$)} & \\
\tableline
NGC 253 & $-$19.95 $\pm$ 0.2\tablenotemark{b} & 1.3 $\pm$ 0.5
	& 0.49 & 8.1 & 6.1 $\pm$ 1.2 & 5,7,27,33,34,35 \\
NGC 628	& $-$20.84 $\pm$ 0.10 & 8.2 $\pm$ 0.8	
	& 1.5	& 3.5	& \nodata & 1,8,19,27	\\
NGC 891 & $-$20.26 $\pm$ 0.18 & $<$ 0.59 (3$\sigma$)
	& 0.35 & 10.6 & \nodata  & 1,9,27  \\
NGC 925 & $-$19.88 $\pm$ 0.12 & 7.5 $\pm$ 1.5
	& 0.75 & 1.00 & 2.4 $\pm$ 0.1 & 1,19,10,27,36 \\
NGC 1068 & $-$22.14 $\pm$ 0.10 & 9.5 $\pm$ 0.6
	& 10.2 & 77 & \nodata & 1,8,28 \\
NGC 1097 & $-$21.62 $\pm$ 0.07 & 9.3 $\pm$ 1.9
	& 0.83 & 16.9 & \nodata & 1,22,27 \\
NGC 1291 & $-$21.25 $\pm$ 0.04 & 0.08 $\pm$ 0.02 
	& 0.046 & 0.38 & \nodata & 1,11,27 \\
NGC 1313 & $-$19.61 $\pm$ 0.2 & 3.6 $\pm$ 0.4
	& 0.47 & 0.70 & \nodata & 1,23,27 \\
NGC 1317 & $-$20.55 $\pm$ 0.06 & 0.86 $\pm$ 0.06 
	& 0.26 & 2.04 & \nodata & 1,12,28 \\
NGC 1365 & $-$22.00 $\pm$ 0.07 & 24 $\pm$ 5
	& \nodata & 46 & \nodata & 1,27 \\
NGC 1512 & $-$19.67 $\pm$ 0.10 & 0.85 $\pm$ 0.09
	& \nodata & 0.53 & \nodata & 1,28 \\
NGC 1566 & $-$21.53 $\pm$ 0.03 & 35 $\pm$ 2
	& \nodata & 7.2 & \nodata & 1,28 \\
NGC 1672 & $-$21.33 $\pm$ 0.08 & 7.4 $\pm$ 0.7
	& \nodata & 9.6 & \nodata & 1,28 \\
NGC 2146 & $-$20.54 $\pm$ 0.13 & $<$ 2.6 (3$\sigma$)
	& 1.14 & 27.9 & 46 $\pm$ 13 & 1,19,28,37,35 \\
NGC 2403 & $-$19.06 $\pm$ 0.08 & 1.89 $\pm$ 0.11
	& 0.49 & 0.72 & \nodata & 1,10,27 \\
NGC 2551 & $-$20.43 $\pm$ 0.20 & 0.5 $\pm$ 0.25
	& \nodata & 0.41 & \nodata & 1,28 \\
NGC 2841 & $-$20.60 $\pm$ 0.10 & 1.30 $\pm$ 0.25 
	& 0.20 & 1.13 & \nodata & 1,19,10,27 \\
NGC 2903 & $-$20.09 $\pm$ 0.10 & 1.6 $\pm$ 0.1
	& 0.74 & 3.4 & 4.5 $\pm$ 1.2 & 1,19,10,27,37,35 \\
NGC 2993 & $-$20.09 $\pm$ 0.14 & 10.2 $\pm$ 0.4
	& 2.7 & 12.3 & \nodata & 1,24,28 \\
NGC 3031 & $-$20.92 $\pm$ 0.03 & 1.7 $\pm$ 0.3 
	& 0.45	& 1.04 & 0.5 $\pm$ 0.2 & 1,10,27,37,35 \\
NGC 3034 & $-$19.72 $\pm$ 0.09 & 0.19 $\pm$ 0.07
	& 1.3	& 14.6	& 10 $\pm$ 3 & 1,19,27,38 \\
NGC 3310 & $-$19.99 $\pm$ 0.10 & 11.9 $\pm$ 0.4
	& 3.0 & 6.9 & 4.6 $\pm$ 3.5 & 1,8,19,29,37,35 \\
NGC 3351 & $-$20.13 $\pm$ 0.10  & 0.72 $\pm$ 0.11
	& 0.47 & 1.60 & \nodata & 1,8,19,28 \\
NGC 3389 & $-$20.09 $\pm$ 0.06 & 5.3 $\pm$ 0.1
	& 1.08 & 3.0 & \nodata & 1,8,30 \\
NGC 4038/9 & $-$21.4 $\pm$ 0.2 & 15.1 $\pm$ 0.5
	& 6.4 & 21.1 & 51 $\pm$ 7 & 2,19,13,29,39,35 \\
NGC 4156 & $-$21.70 $\pm$ 0.05 & 6.5 $\pm$ 1.5
	& \nodata & \nodata & \nodata & 1 \\
NGC 4214 & $-$18.41 $\pm$ 0.15 & 2.03 $\pm$ 0.08
	& 0.30 & 0.33 & \nodata & 1,19,14,29 \\
NGC 4258 & $-$20.81 $\pm$ 0.08 & 2.7 $\pm$ 0.5
	& 1.3 & 1.96 & \nodata & 1,10,27 \\
NGC 4321 & $-$21.76 $\pm$ 0.08 & 9.4 $\pm$ 2.8 
	& 2.2 & 10.3 & \nodata & 1,19,8,29 \\
NGC 4449 & $-$18.14 $\pm$ 0.13  & 2.14 $\pm$ 0.05
	& 0.35 & 0.47 & 0.71 $\pm$ 0.08 & 1,8,14,31,40,35 \\
NGC 4631 & $-$20.49 $\pm$ 0.16 & 10.2 $\pm$ 0.6 
	& 1.83 & 8.2 & 3.6 $\pm$ 2.4 & 1,8,9,27,37,35 \\
NGC 4647 & $-$20.30 $\pm$ 0.08 & 2.1 $\pm$ 0.2
	& 0.74 & 4.1 & \nodata & 1,19,13,29 \\
NGC 4736 & $-$20.22 $\pm$ 0.13 & 1.79 $\pm$ 0.05 
	& 0.58 & 1.87 & 1.6 $\pm$ 0.4 & 1,19,8,27,37,35 \\
NGC 5055 & $-$20.87 $\pm$ 0.10 & 2.9 $\pm$ 0.3
	& 0.88  & 5.2 & 3.8 $\pm$ 1.3 & 1,8,27,37,35  \\
NGC 5194 & $-$21.07 $\pm$ 0.06 & 9.8 $\pm$ 0.4 
	& 2.2	& 9.4 & 5.5 $\pm$ 1.1 & 1,19,8,40,27,37,35 \\
NGC 5236 & $-$20.52 $\pm$ 0.04 & 5.3 $\pm$ 0.2
	& 2.0 & 5.6 & \nodata   & 1,10,27  \\
NGC 5253 & $-$17.53 $\pm$ 0.12 & 0.88 $\pm$ 0.09
	& 0.30 & 0.53 & \nodata & 1,21,14,28 \\
NGC 5457 & $-$21.51 $\pm$ 0.10 & 23.1 $\pm$ 0.5
	& 3.1	& 7.6 & 3.4 $\pm$ 1.3   & 1,10,27,37,35 \\
NGC 6090 & $-$21.34 $\pm$ 0.1 & 22.8 $\pm$ 0.7
	& 11.7 & 96 & \nodata & 3,20,29 \\
NGC 6946 & $-$21.31 $\pm$ 0.11 & $<$ 18 (3$\sigma$)
	& 2.6 & 8.0 & 5.1 $\pm$ 0.9 & 1,19,10,27,37,35 \\
DDO 50	& $-$16.90 $\pm$ 0.17	& 0.58 $\pm$ 0.03
	& 0.044 & 0.011 & 0.032 $\pm$ 0.005 & 1,21,15,28,41 \\
DDO 66	& $-$13.76 $\pm$ 0.31	& 0.017 $\pm$ 0.003 
	& 0.0016 & $<$ 0.0050 & \nodata	& 1,21,15,32 \\
DDO 75  & $-$14.48 $\pm$ 0.11	& 0.085 $\pm$ 0.008
	& 0.012 & 0.00061 & \nodata  & 1,16,28 \\
DDO 81  & $-$17.22 $\pm$ 0.21 & 1.00 $\pm$ 0.15 
	& 0.076 & 0.043 & \nodata & 1,15,27 \\
UGC 6697 & $-$21.25 $\pm$ 0.10 & 35 $\pm$ 5
	& 5.3 & 11.0 & \nodata & 1,17,18,28 \\
UGC 7188 & $\sim$$-$16.4 $\pm$ 0.7\tablenotemark{c} & 0.14 $\pm$ 0.03
	& \nodata & \nodata & \nodata & 21 \\
CGCG 097-093 & $-$19.17 $\pm$ 0.2 & 2.6 $\pm$ 0.3
	& 0.27 & \nodata & \nodata & 1,25,18 \\
CGCG 097-114 & $-$20.13 $\pm$ 0.10 & 7 $\pm$ 2
	& 2.2 $\pm$ 0.5 & \nodata & \nodata & 2,25 \\
Mrk 66	& $-$19.75 $\pm$ 0.1 & 16 $\pm$ 1
	& 1.1 & 2.1 & \nodata & 3,26,28 \\
IRAS 08339+6517 & 
	$\sim$$-$21.2 $\pm$ 0.3\tablenotemark{a} & 54 $\pm$ 3
	& 11.2	& 36 & \nodata & 4,20,29 \\
\tableline \\
\end{tabular}
\end{center}
\tablerefs{ 
	(1) \cite{rc3} (2) \cite{dev88} (3) \cite{huchra77}
	(4) \cite{kirshner78} (5) \cite{esolv} (6) \cite{garnier96} 
	(7) \cite{hoopes96} (8) \cite{kk83} times 1.17
	(9) \cite{hoopes99} (10) Kennicutt (unpublished) 
	(11) \cite{caldwell91} (12) \cite{crocker96} (13) \cite{k87} 
	(14) \cite{martin98} (15) \cite{miller94} (16) \cite{hunter96} 
	(17) \cite{k84} (18) \cite{gavazzi98} (19) \cite{young96} 
	(20) \cite{gonzalez98} (21) \cite{iglesias99} (22) \cite{hummel87} 
	(23) \cite{ryder94} (24) \cite{usui98} (25) \cite{moss98} 
	(26) \cite{mcquade95} (27) \cite{rice88} (28) \cite{moshir90} 
	(29) \cite{soifer89} (30) \cite{bushouse88} (31) \cite{young89}
	(32) \cite{melisse94} (33) \cite{wright90} (34) \cite{kuhr81} 
	(35) \cite{niklas97} (36) \cite{duric88} (37) \cite{white92}
	(38) \cite{carlstrom91} (39) \cite{condon90} (40) \cite{klein96}
	(41) \cite{tongue95} }
\tablenotetext{a}{Interpolated between photographic $J$ and $F$ data: 
	\citet{kirshner78} }
\tablenotetext{b}{Assuming $B - V \sim 1$ from $B - R = 1.52$: \citet{esolv}}
\tablenotetext{c}{Assuming $B - V \sim 0.5$: \citet{garnier96}}
\end{scriptsize}
\end{table}

\end{document}